\documentclass[aps,prl,reprint,groupedaddress,showpacs,superscriptaddress]{revtex4-1}
\usepackage{amsmath,amssymb,amsthm,color}
\usepackage{graphicx}
\usepackage{subfigure}
\usepackage{ulem}
\usepackage{hyperref}

\begin{document}

\title{Non-Abelian Evolution of a Majorana Train in a Single Josephson Junction}

\author{Sang-Jun Choi}
\affiliation{Department of Physics, Korea Advanced Institute of Science and Technology, Daejeon 34141, Korea}
\affiliation{\mbox{Center for Theoretical Physics of Complex Systems, Institute for Basic Science, Daejeon 34126, Korea}}
\author{H.-S. Sim} \email[Corresponding author.]{hssim@kaist.ac.kr}
\affiliation{Department of Physics, Korea Advanced Institute of Science and Technology, Daejeon 34141, Korea}
\date{\today}

\begin{abstract} 
Demonstration of non-Abelian anyon statistics often requires dynamical controls of a complicated device that are challenging in realistic situations.
We propose a {\it single} Josephson junction to detect a non-Abelian statistics effect of Majorana fermions, formed by two finite-size $s$-wave superconductors on a topological insulator under a magnetic field. At certain field strengths, a train of three localized Majorana fermions appears along the junction, while an extended chiral Majorana fermion encircles the train and the superconductors. A DC voltage bias across the junction causes the train to move and collide with the extended Majorana fermion. This involves interchange of fusion partners among the four Majorana fermions.
This gives rise to non-Abelian state evolution and a $2n\pi$ fractional AC Josephson effect. The period-elongation factor $n$ is an integer $n\ge2$ tunable by the voltage.
\end{abstract}
   



\maketitle

When non-Abelian anyons adiabatically exchange their positions and fuse, the system can evolve from one state to another~\cite{Nayak,Stern, Alicea, Flensberg,Aguado}. The non-Abelian statistics is a key of topological quantum computing where one operates qubits in a nonlocal way immune to local decoherence. 

There are efforts to realize non-Abelian anyons. For example, Majorana fermions emerge in effective $p$-wave superconductors (SCs)~\cite{Lutchyn, Oreg, FuKane08, Yazdani, Qi, ShtengelPRX}. Their existence is supported by observation of zero-bias conductance peaks~\cite{Kouwenhoven, Heiblum, Deng, YazdaniExp, Jin, Sun, Hao} and fractional AC Josephson effects~\cite{Furdyna, Wiedenmann,Deacon,Laroche}. Their non-Abelian statistics, though, has not yet been detected. Related proposals utilize Josephson tri-junctions on a topological insulator~\cite{FuKane08} or nanowire junctions~\cite{Fisher,Aasen,Beenakker1, Beenakker2}.
These require dynamical control of a number of electrical gates or SC phase differences in multiple junctions, and need additional setups to detect the statistics. 
  
 
Fractional Josephson effects are hallmarks of topological SCs. A Majorana zero mode (MZM) allows single-electron transfer across a Josephson junction, resulting in $4 \pi$ periodic current as a function of the SC phase difference of the junction~\cite{FuKanePRB,Kitaev1D,Kwon}.
Fractional Josephson effects of a longer period can occur when electron-electron interactions or parafermions play a role~\cite{Clarke,ZhangKane,Orth,Klinovaja,Peng,Zazunov}.
The present work predicts a new fractional Josephson effect originating from the non-Abelian statistics.

\begin{figure}[b]
\centering
\includegraphics[width=\columnwidth]{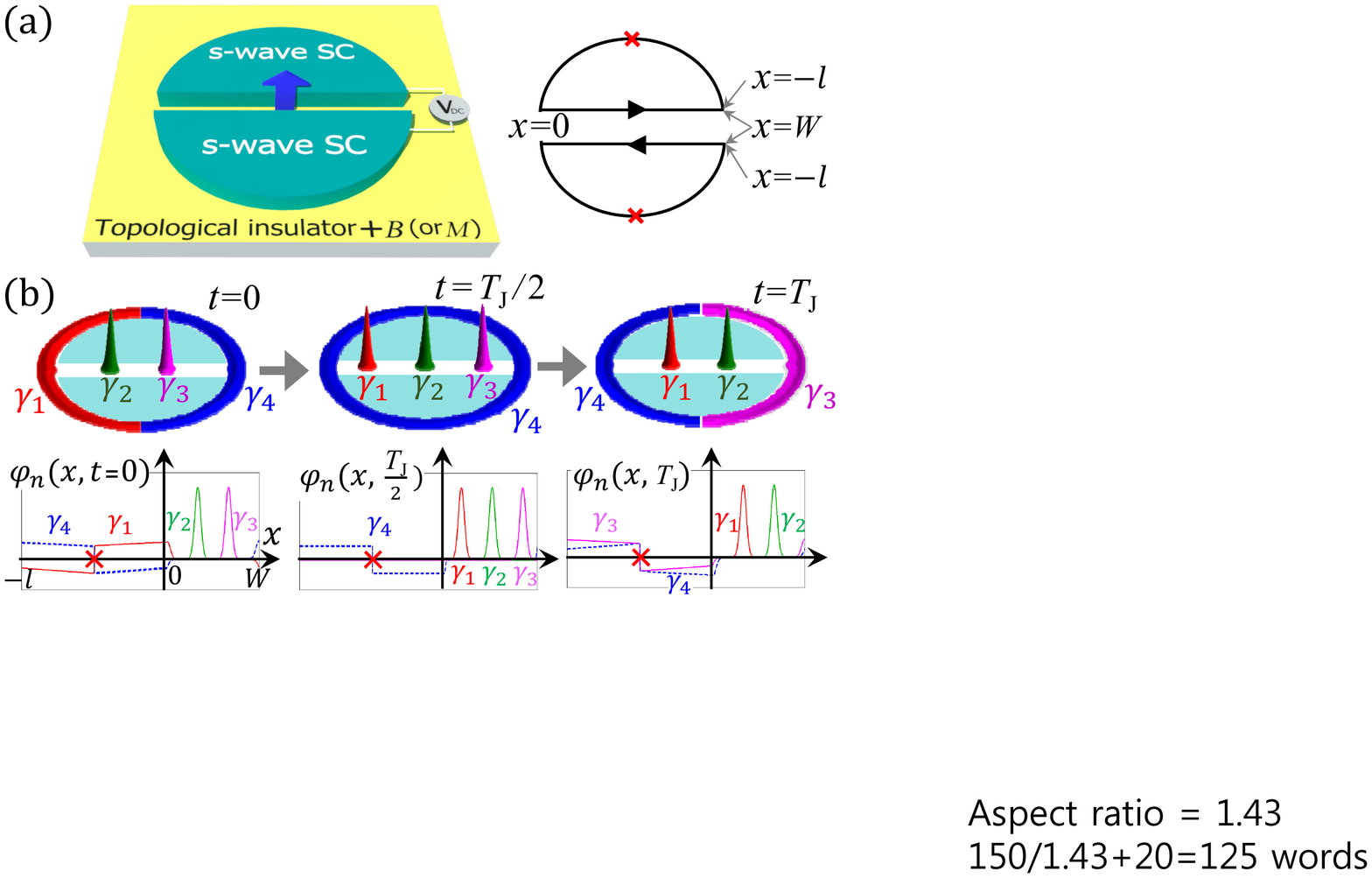}
\caption{(a)  Topological Josephson junction under magnetic fields $B$. 
Right: Coordinate $x \in [-l,W]$ of the SC edges, along which chiral Majorana modes propagate (arrows).  
Crosses depict branch cuts at $x=-l/2$ on the arcs. 
(b) When three magnetic flux quanta (thick arrow) pierce the junction, four Majorana fermions $\gamma_{k=1,2,3,4}$ appear.  Their probability distribution $|\varphi_k(x,t)|^2$ (upper panels) and wave function $\varphi_k(x,t)$ (lower) are numerically obtained. At time $t=0$, $\gamma_1$ and $\gamma_4$ extend and overlap along the arcs, forming a fusion pair. $\gamma_2$ and $\gamma_3$ are localized MZMs in the junction and compose another fusion pair.
Voltage bias $V_\textrm{DC}$ across the junction makes $\gamma_{k}$'s move.
At $t = T_\text{J}/2$, all $\gamma_k$'s become MZMs.
At $T_\text{J}$, $\gamma_3$ collides with the arcs, fusing with $\gamma_4$, while $\gamma_1$ fuses with $\gamma_2$. 
The fusion partners interchange as $\{(\gamma_4,\gamma_1), (\gamma_3,\gamma_2)\}_{t=0}\to\{(\gamma_4,\gamma_3), (\gamma_2,\gamma_1)\}_{T_\textrm{J}}$, leading to non-Abelian state evolution.
} 
\label{Fig:setup}
\end{figure}

In this work, we propose a {\it single} Josephson junction to detect the non-Abelian statistics.
It hosts four Majorana fermions $\gamma_{k=1,2,3,4}$'s with the help~\cite{twoMFs} of an external magnetic field  (Fig.~\ref{Fig:setup}).
Voltage bias $V_\textrm{DC}$ across the junction makes $\gamma_k$'s move and collide each other.
This involves interchange of fusion partners among $\gamma_k$'s, causing non-Abelian state evolution, such as $|00\rangle\mapsto\left(e^{i \phi_+}|00 \rangle-ie^{i \phi_-}|11 \rangle\right)/\sqrt{2}$,  in one conventional Josephson period $T_\textrm{J}=h/(2eV_\textrm{DC})$. $|00\rangle$ and $|11\rangle$ are fermion occupation states formed by $\gamma_k$'s. $\phi_\pm$'s are $V_\textrm{DC}$-dependent dynamical phases arising from fusion and splitting of $\gamma_{k}$'s.
The evolution returns to the initial state after time $n T_\textrm{J}$, showing a $2 n \pi$ fractional AC Josephson effect. 
The period-elongation factor $n$ is an integer $\ge 2$ tunable by $V_\textrm{DC}$. The effect can be detected by Shapiro spikes.

{\it Setup.---} Figure~\ref{Fig:setup}(a) shows two topological SCs induced by $s$-wave SCs on a topological insulator (TI) surface~\cite{Williams}. Perpendicular magnetic fields $B$ (or Zeeman fields $M$ by magnetic insulators), applied to the surface outside the SCs, break the time reversal symmetry,  opening an energy gap at the Fermi energy.
Then an extended chiral Majorana mode u (l), whose operator is $\eta_\textrm{u(l)}(x)=\eta_\textrm{u(l)}^\dagger(x)$, is formed along the boundary between the upper (lower) topological SC and the gapped TI region due to topological origin~\cite{FuKane08, Bruder, HSSim, FuKane09, Akhmerov, Law}. $x\in[-l,W]$ is the coordinate along the SC edge.
The mode gains Berry phase $\pi$ after one circulation along the edge, when each SC has no vortex.
To describe this, we place a branch cut~\cite{HSSim}   
at $x =-l/2$:
$\eta_\textrm{u(l)}(-l/2+0^+) = -\eta_\textrm{u(l)}(-l/2-0^+)$ with positive infinitesimal $0^+$,
and choose $\eta_\textrm{u(l)}(-l)=\eta_\textrm{u(l)}(W)$.


The two SCs form a short Josephson junction of length $L$ and width $W$. 
Three magnetic flux quanta are enclosed by the junction area, $N=BLW/\Phi_0=3$, and $e$ is the electron charge. 
The junction Hamiltonian~\cite{FuKane08,ChoiSimPRB} is 
\begin{equation} \label{originalH}
H(t)=\int_{-l}^{W} dx \Gamma(x)^\top\left( -i\hbar v(x) \sigma_z \partial_x  + m(x,t) \sigma_y \right)\Gamma(x). \end{equation}
$\sigma_{x,y,z}$ are Pauli matrices and $\Gamma(x)^\top \equiv \left(\eta_\textrm{u}(x),\eta_\textrm{l}(x) \right)$.
Inside the junction $x\in[0,W]$, the modes $\eta_\textrm{u}$ and $\eta_\textrm{l}$ counterpropagate with velocity $v(x)=v_\textrm{J}$ and couple each other with strength $m(x,t) = \Delta_0 \sin\left(\frac{N \pi x}{W} - \frac{eV_\textrm{DC}t}{\hbar} \right)$.
$\Delta_0$ is the gap of the topological SCs.
$m(x,t)$ depends on $x$ and time $t$, since the voltage $V_\textrm{DC}$ and the magnetic field affect the SC phase difference of the junction. Along the arcs $x\in [-l,0]$ of the SCs, $\eta$'s propagate with velocity $v(x)=v_\textrm{arc}$ and $m(x,t)=0$.
The upper and lower arcs have the same length $l$ for simplicity; our results are qualitatively unchanged when their lengths are different. 

\begin{figure}[tb]
\centering
\includegraphics[width=\columnwidth]{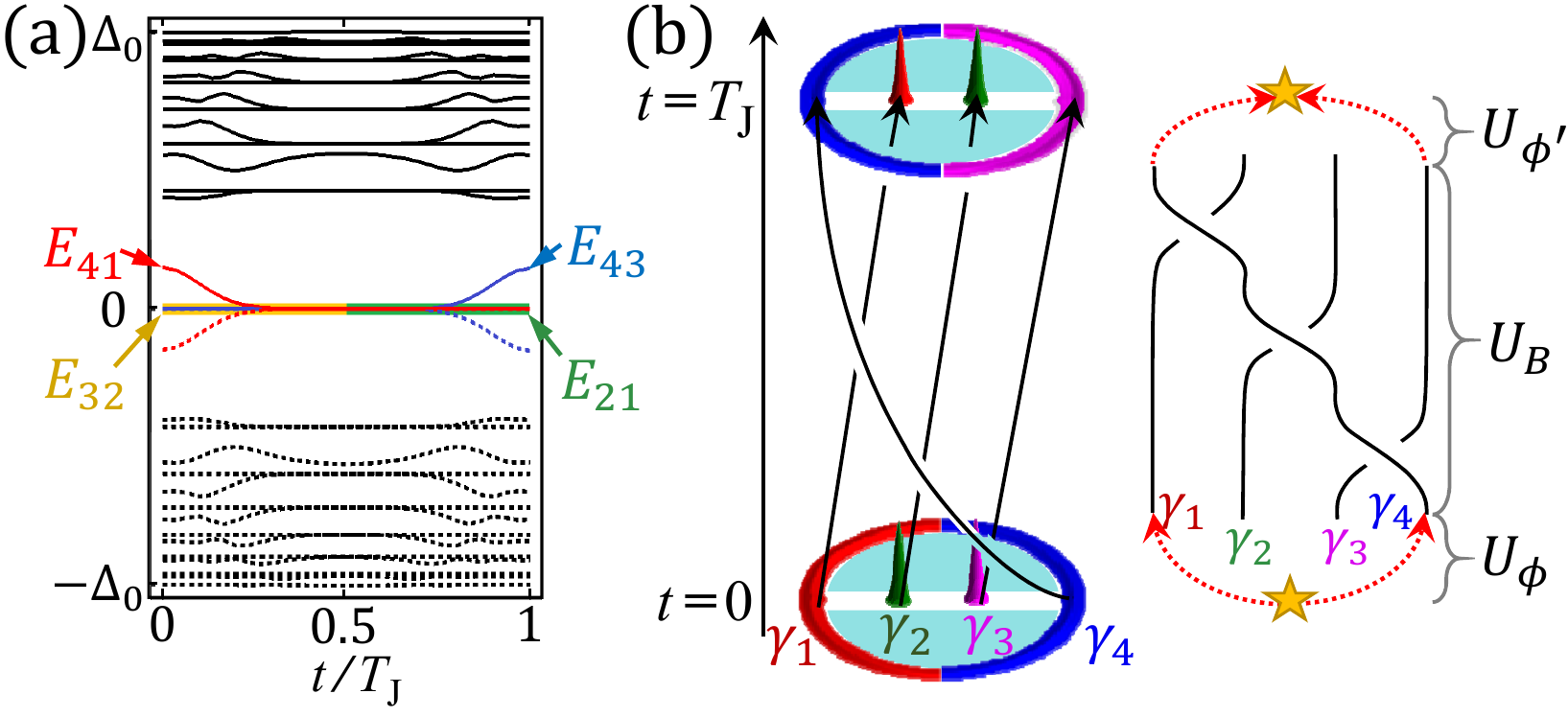}
\caption{(a) Numerical results of single-particle energy levels of the setup.
Energy levels $E_{kk'}$ formed by the Majorana fermions $\gamma_{k=1,2,3,4}$ are marked. 
We choose the SC gap $\Delta_0 \sim 1 \,\text{meV}$ (proximity-induced, e.g, by NbN SCs~\cite{Lin}), SC coherence length $\xi = \hbar v_F / \Delta_0 \sim 260 \, \textrm{nm}$ (Fermi velocity $v_F \sim 4.0 \times 10^5 \,\text{m/s}$~\cite{Qu}),  a short junction of $L \sim 60 \,\text{nm} < \xi$ and $W \sim 800 \,\text{nm}$, $B \sim 130 \,\text{mT}$ ($N=3$), $2l \sim 2.5\,\mu\text{m}$, $v_\textrm{arc} \sim v_F$, $v_\text{J} \sim 0.04 \, v_F$~\cite{ChoiSimPRB}, and $M_\textrm{u}=M_\textrm{l}=0$. These parameters satisfy $\lambda \ll W/3$.
(b) World lines equivalent to the evolution of $\gamma_k$'s during one Hamiltonian period $T_\text{J}$.
The fusion of $\gamma_4(t)$ and $\gamma_1(t)$ in $t \in [0,T_\textrm{J}/2]$ and that of $\gamma_4(t)$ and $\gamma_3(t)$ in $[T_\textrm{J}/2,T_\textrm{J}]$ are depicted by stars.} 
\label{Fig:energy}
\end{figure}

{\it Majorana train.---}  We write the Hamiltonian as $H(t)=i \sum_{q=1,2,\cdots}E_q(t)\gamma_{2q}(t)\gamma_{2q-1}(t)$, using its particle-hole symmetry. 
The Majorana operators $\gamma_{j = 2q-1, 2q}$ associated with the single-particle level $E_q$ are found as $\gamma_{j}(t)=\int^W_{-l}dx[\eta_\textrm{u}(x)-(-1)^j\eta_\textrm{l}(x)]\varphi_j(x,t)$. $\varphi_j(x,t)$ is the real wave function of $\gamma_j$.
We obtain the levels $E_q(t)$ in Fig.~\ref{Fig:energy}(a), numerically solving Eq.~\eqref{originalH} with realistic parameters that satisfy the conditions ($\lambda \ll W/3$, $T_J \gg \hbar/E_0$) explained later.
The two lowest levels have zero energy at certain times.
We focus on the four Majorana fermions $\gamma_{k=1,2,3,4}$ associated with the two levels, labeling them with another index $k$.
Figure~\ref{Fig:setup}(b) shows their wave functions.
  
Along the junction, a train of localized Majorana fermions $\gamma_k(t)$ appears
at positions
\begin{equation}\label{positionMZM}
x_{k = 1,2,\cdots}(t) = \frac{W}{N} \left(k - 1 + \frac{t}{T_\text{J}}\right) \quad \in [0,W], 
\end{equation}
at which the SC phase difference is an integer multiple of $\pi$ and $m(x,t)$ has sign change~\cite{twoMFs}.
Because of $V_\textrm{DC}$ the train moves with periodic position shift $x_k (T_\textrm{J})=x_{k+1}(0)$.
Distance between adjacent $\gamma_k$'s is $W/N = W/3$.
A Majorana fermion $\gamma_k$ is a localized MZM (well separated from its neighbors $\gamma_{k\pm1}$ and the SC arcs) in a Gaussian wave function~\cite{ChoiSimPRB} with localization length $\lambda=\sqrt{{\hbar v_\text{J} W}/({N \pi \Delta_0})}$,
when  $\lambda \ll W/3, x_k, W-x_k$.



Along the arcs, extended chiral Majorana fermions occur. We find their energy $E$ quantization condition~\cite{ChoiSimPRB},
\begin{equation} \label{quantization}
2l E/(\hbar v_\textrm{arc}) + \pi  + \pi (M_\textrm{J} + M_\textrm{u} + M_\textrm{l}) = 0, \pm 2\pi, \cdots.
\end{equation}
The first two terms are the dynamical phase and Berry phase gained in one circulation along the upper and lower arcs of total length $2l$.
$M_\textrm{u(l)}$ is the number of MZMs in vortices inside the upper (lower) SC.
$M_\textrm{J}$ is the number of the localized MZMs [equivalently, the number of sign changes of $m(x,t)$ in Eq.~\eqref{positionMZM}] inside the junction~\cite{twoMFs}. The condition implies that an extended chiral MZM appears along the arcs, when the total number of MZMs inside the junction and the two SCs is odd. The appearance is insensitive to the details of $m(x,t)$ and the arcs. 
This chiral MZM was not considered in previous works~\cite{twoMFs}.
We will consider the case of $M_\textrm{u} = M_\textrm{l} = 0$~\cite{ChoiSimPRB}.
 

{\it Fusion-partner interchange.---}  When two MZMs $\gamma_k$ and $\gamma_{k'}$ collide,
they fuse to form
fermion occupation states $|{0}_{kk'}\rangle_t$, $|{1}_{kk'} \rangle_t \!=\! f^\dagger_{kk'} (t) |0_{kk'} \rangle_t$. Here $f_{kk'}(t) \!=\! [\gamma_{k}(t) + i\gamma_{k'}(t)]/2$. 
As shown below, a sequence of such fusions occur in our system, resulting in non-Abelian state evolution in the 
adiabatic regime of $T_J \gg \hbar/E_0$ that the energy gap $E_0$ between the two lowest levels by $\gamma_k$'s and
the other midgap levels is much larger than $2e V_\textrm{DC}$.

At $t=0$, the Majorana fermions $\gamma_{k=1,2,3,4}$'s appear at $x=(k-1)W/3$ along the junction [see Fig.~\ref{Fig:setup}(b)].
$\gamma_1$ and $\gamma_4$ are localized in the junction but overlap each other along the arcs, fusing into 
a state $|0_{41} \rangle_{t=0}$ of energy $- E_{41}(t=0)<0$ and  $|1_{41} \rangle_{t=0}$ of $E_{41}(t=0)$.
When $\lambda \ll W/3$, $\gamma_2$ and $\gamma_3$ are localized MZMs and form fusion states $|0_{32} \rangle_{t=0}$ and $|1_{32} \rangle_{t=0}$ with zero energy $E_{32}=0$.
Hence the junction has two-fold degenerate ground states, $|0_{41}0_{32}\cdots\rangle_0$ and $|0_{41}1_{32}\cdots\rangle_0$~\cite{stateR}.
We follow the time evolution of  the even-parity ground state $|\psi(t=0)\rangle = |0_{41}0_{32}\rangle_0$. 
Here the part $|\cdots\rangle_0$ for the other midgap levels is ignored, since its evolution is trivial~\cite{ChoiSimPRB}.

In $t\in[0,T_\textrm{J}/2]$, $\gamma_{1,2,3}$'s move along the junction and $\gamma_4$ further moves into the arcs.
$\gamma_2(t)$ and $\gamma_3(t)$ remain MZMs ($E_{32}=0$), while
$\gamma_4(t)$ and $\gamma_1(t)$ become split and their energy $E_{41}(t)$ decreases with $t$.
At $t=T_\textrm{J}/2$, $\gamma_{1,2,3,4}(t)$'s all become MZMs, having no overlap each other.
$\gamma_4(T_\textrm{J}/2)$ is the extended chiral MZM along the arcs, obeying Eq.~\eqref{quantization} at $E=0$.
Hence the low-energy part of $H(t)$ is reduced into the Hamiltonian $H_\textrm{MF}(t)=i E_{41}(t) \gamma_4(t) \gamma_1(t) = E_{41}(t)[2f^{\dagger}_{41}(t)f_{41}(t)-1]$. Notice $E_{41}(T_\textrm{J}/2)=0$.
The initial state $|0_{41}0_{32}\rangle_0$ adiabatically evolves into $|\psi(t=T_\textrm{J}/2)\rangle=e^{i\phi}|0_{41}0_{32}\rangle_{T_\textrm{J}/2}$,
gaining dynamical phase $\phi = \int_0^{T_\textrm{J}/2} dt' E_{41}(t')/\hbar$.
  

In $t\in[T_\textrm{J}/2,T_\textrm{J}]$, $\gamma_3(t)$ approaches the arcs and fuses with $\gamma_4(t)$, forming states $|0_{43}\rangle_t$ and $|1_{43}\rangle_t$ of energy $\mp E_{43}(t)$. $E_{43}(t)$ increases with $t$. $\gamma_1(t)$ and $\gamma_2(t)$ are MZMs, and their zero-energy fusion states are $|0_{21}\rangle_t$ and $|1_{21}\rangle_t$.
Hence $H_\textrm{MF}(t)= E_{43}(t)[2f^{\dagger}_{43}(t)f_{43}(t)-1]$ in this domain.
The fusion-partner interchange happens as $\{(\gamma_4,\gamma_1), (\gamma_3,\gamma_2)\}_{t \in [0,T_\textrm{J}/2]} \to \{(\gamma_4,\gamma_3), (\gamma_2,\gamma_1)\}_{[T_\textrm{J}/2,T_\textrm{J}]}$.
To find $|\psi(t)\rangle$ in $t\in[T_\textrm{J}/2,T_\textrm{J}]$, we write $|\psi(t=T_\textrm{J}/2)\rangle=e^{i\phi}|0_{41}0_{32}\rangle_{T_\textrm{J}/2}$ in terms of the new fusion states,
\begin{eqnarray}
e^{i\phi}|0_{41}0_{32}\rangle_{T_\textrm{J}/2} = \frac{e^{i \phi}}{\sqrt{2}}(|0_{43}0_{21} \rangle_{T_\textrm{J}/2}+|1_{43}1_{21} \rangle_{T_\textrm{J}/2}), \label{halfperiod}
\end{eqnarray}
in accord with the fusion rule $\sigma\times\sigma=I+\psi$~\cite{Nayak} 
of $\gamma_k(T_\textrm{J}/2)$'s.
The state adiabatically evolves into $|\psi(T_\text{J})\rangle = \frac{e^{i\phi}}{\sqrt{2}}(e^{i\phi'}|0_{43}0_{21}\rangle_{T_\text{J}} + e^{-i\phi'}|1_{43}1_{21}\rangle_{T_\text{J}})$ at $t=T_\textrm{J}$, as $|0_{43}0_{21} \rangle_t$ and $|1_{43}1_{21} \rangle_t$ are the eigenstates of $H_\textrm{MF}(t)$ in $t \in [T_\textrm{J}/2,T_\textrm{J}]$. The dynamical phase $\pm \phi'= \mp \int_{T_\text{J}/2}^{T_\text{J}} dt' E_{43}(t') / \hbar$ arises by
the fusion of $\gamma_3$ and $\gamma_4$.
 
 
{\it Non-Abelian evolution.---} To see the non-Abelian nature of the state evolution, we find the relations between the wave functions of $\gamma_k$'s at $t=0$ and $T_\textrm{J}$  in Fig.~\ref{Fig:setup}(b),
\begin{equation} 
\varphi_{k=1,2,3} (x,T_\text{J}) \!=\! \varphi_{k+1} (x,0), \,\,\, \varphi_4 (x,T_\text{J}) \!=\! - \varphi_1 (x,0), \label{evolutionMF}
\end{equation}
The factor $-1$ accords with the phase gain of a Majorana vortex that exchanges positions with three other vortices, passing their branch cuts.
Equivalently,
$\gamma_{k=1,2,3}(T_\text{J}) = \gamma_{k+1}(0)$, $\gamma_4(T_\textrm{J}) = -\gamma_1(0)$, $f_{21}(T_\textrm{J}) = f_{32}(0)$, $f_{43}(T_\textrm{J}) = i f_{41} (0)$.
Using this we find non-Abelian evolution $|\psi(0)\rangle \mapsto |\psi(T_\text{J})\rangle$ in one Hamiltonian period,
\begin{equation}
|0_{41}0_{32}\rangle_{0} \mapsto |\psi(T_\text{J})\rangle = \frac{e^{i\phi}}{\sqrt{2}}(e^{i\phi'}|0_{41}0_{32}\rangle_0 - i e^{-i\phi'}|1_{41}1_{32}\rangle_0). \label{evolution}
\end{equation}

We obtain the non-Abelian evolution of general initial states, $|\psi(0)\rangle \mapsto |\psi(T_\textrm{J})\rangle = U |\psi(0)\rangle$, in the $t=0$ basis $\{| 0_{41}0_{32} \rangle_0, |1_{41}1_{32}\rangle_0, |0_{41}1_{32}\rangle_0, |1_{41}0_{32} \rangle_0 \}$,
\begin{equation}\label{U}
U
\!=\!
\frac{1}{\sqrt{2}}
\left(
{\setlength\arraycolsep{1.7pt}
\begin{array}{cccc}
e^{i\phi_+} & i e^{-i\phi_-}  & 0 & 0 \\
-i e^{i\phi_-} & -e^{-i\phi_+} & 0 & 0 \\
0 & 0 &  -i e^{i\phi_+}  & e^{-i\phi_-} \\
0 & 0 & e^{i\phi_-}  &  -i e^{-i\phi_+}
\end{array}
}
\right)
\!= \!
\left(
{\setlength\arraycolsep{1pt}
\begin{array}{cc}
U_e & \bf0 \\
\bf0 & U_o
\end{array}
}
\right). 
\end{equation}
Here, $\phi_\pm =  \phi \pm \phi'$. 
It is decomposed as $U = U_{\phi'} U_\textrm{B} U_\phi$,
\begin{eqnarray}
U_{\phi}
=
\left(
{\setlength\arraycolsep{1pt}
\begin{array}{cccc}
e^{i\phi} & 0 & 0 & 0 \\
0 & e^{-i\phi} & 0 & 0 \\
0 & 0 & e^{i\phi} & 0 \\
0 & 0 & 0 & e^{-i\phi} 
\end{array}
}
\right),
\,\,\, &
U_\textrm{B}	
=\frac{1}{\sqrt{2}}
\left(
{\setlength\arraycolsep{2.5pt}
\begin{array}{cccc}
1 & i & 0 & 0 \\
-i & -1 & 0 & 0 \\
0 & 0 & -i & 1 \\
0 & 0 & 1 & -i
\end{array}
}
\right). \nonumber
\end{eqnarray}
$U_{\phi}$ describes dynamical phase gain by the splitting of $\gamma_k$'s in $t\in[0,T_\textrm{J}/2]$, while 
$U_{\phi'}$ is due to the fusion in $[T_\textrm{J}/2,T_\textrm{J}]$.
$U_\textrm{B}$ describes the fusion-partner interchange, 
$\gamma_{k =1,2,3,4}(T_\textrm{J}) = U_\textrm{B} \gamma_k (0) U_\textrm{B}^\dagger$, which agrees Eq.~\eqref{evolutionMF}. 
$U_\textrm{B}$ is further decomposed, $U_\textrm{B} = U_{21} U_{32} U_{43}$, into a series of braidings $U_{ab} = [1 + \gamma_{a} (0) \gamma_{b} (0)]/\sqrt{2}$ of $\gamma_a$ and $\gamma_b$. This coincides with the Ivanov's construction~\cite{Ivanov} obtained from the world lines in Fig.~\ref{Fig:energy}(b).

 
 
From Eq.~\eqref{U} we find that
the evolution $U_e|\varphi_e \rangle$ of an even-parity state $|\varphi_\textrm{e} \rangle$
in one Hamiltonian period $T_\textrm{J}$ is described by its rotation about an axis by rotation angle $\Omega = 2\arccos \left( \frac{\sin\phi_+}{\sqrt{2}} \right)$
on the Bloch sphere for pseudospins $|\mathbf{\Uparrow}\rangle\equiv|0_{41}0_{32}\rangle_0$, $|\mathbf{\Downarrow}\rangle\equiv|1_{41}1_{32}\rangle_0$. 
The state returns to the initial state, up to a phase factor, after $n T_\textrm{J}$.
The period-elongation factor $n$ is the smallest integer satisfying that $n \Omega$ equals a multiple of $2 \pi$. 
Since $\Omega \in [\frac{\pi}{2}, \frac{3\pi}{2}]$ (mod $2\pi$), we find $n\ge2$, reflecting non-Abelian nature.
$n$ is experimentally tunable by $V_\textrm{DC}$, as $\phi_+$ depends on $V_\textrm{DC}$ as $\phi_+=\pi \overline{E}/(eV_\textrm{DC})$ and the average energy change $\overline{E}=[\int_0^{T_\textrm{J}/2} E_{41}(t') dt' + \int_{T_\textrm{J}/2}^{T_\textrm{J}} E_{43}(t') dt']/T_\textrm{J}$ is $V_\textrm{DC}$ independent. Odd-parity states have the same feature.
    
 

\begin{figure}[t]
\centering
\includegraphics[width=0.99\columnwidth]{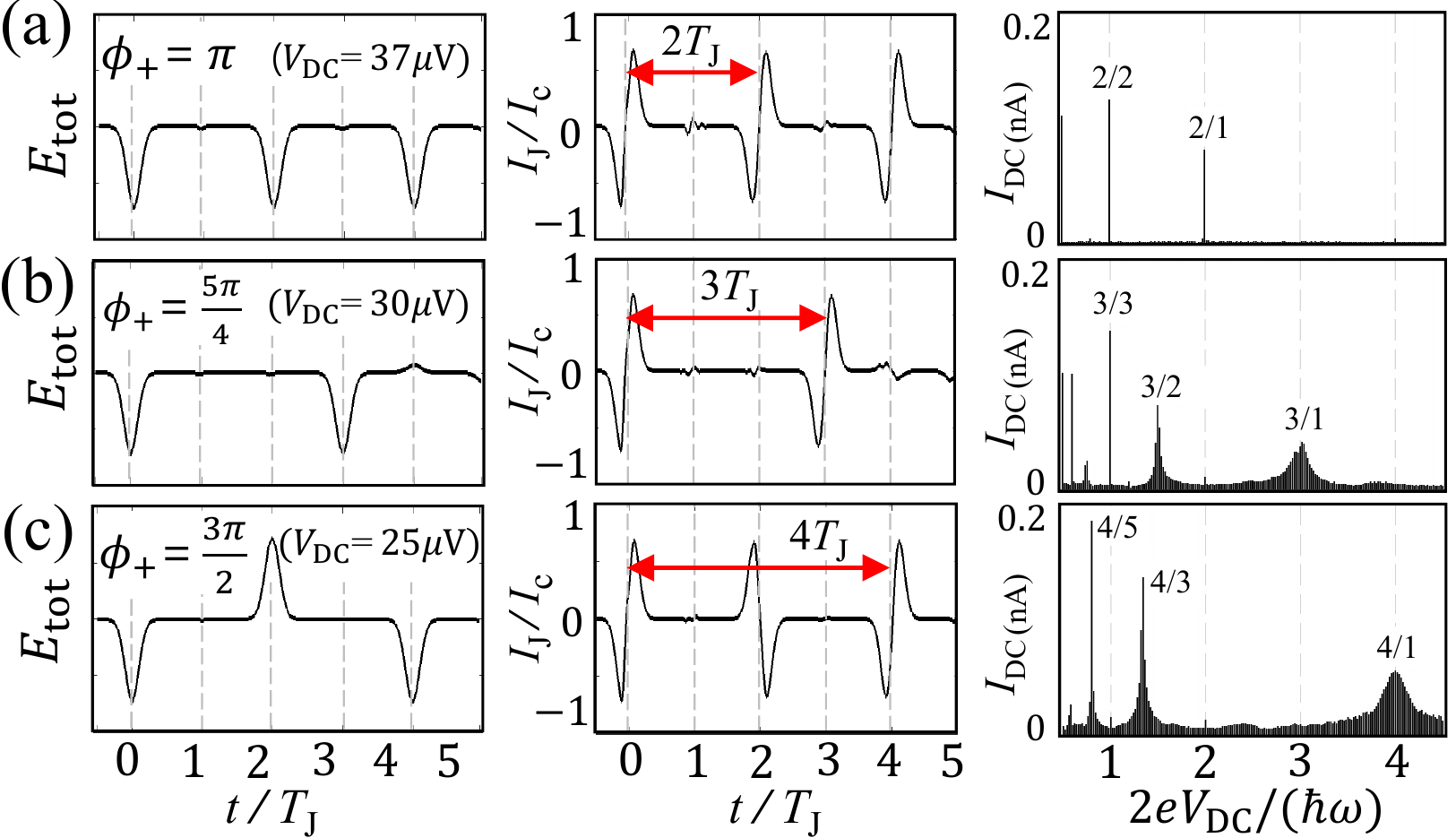}
\caption{$2n\pi$ fractional AC Josephson effect. 
For different $\phi_+$'s and $V_\textrm{DC}$'s, time dependence of the energy $E_\textrm{MF}$ of $|\psi(t)\rangle$ (left panels),  the Josephson current $I_\textrm{MF}$ (normalized by its maximum $I_c$) mediated by $|\psi(t)\rangle$ (middle), 
the Shapiro spikes of long-time average of $I_\textrm{MF}$ measured with an AC voltage of frequency $\omega$ (right). The spike positions $2eV_\textrm{DC} / (\hbar \omega)$ are represented as $n / p$.
$E_\textrm{MF}$ and $I_\textrm{MF}$ have the same period $nT_\text{J}$, (a) $n=2$, (b) $3$, (c) $4$. 
The numerical results are obtained with the initial state $|0_{41}0_{32}\rangle_0$ and the parameters in Fig.~\ref{Fig:energy}. 
}\label{Fig:FJ}
\end{figure}

Figure~\ref{Fig:FJ}(a) shows the energy $E_\textrm{MF}(t)$ of the state $|\psi (t) \rangle$. 
It is $nT_\textrm{J}$ periodic. As an example, we explain the $n=2$ case that happens when $\phi_+ = \pi$ (namely, $\Omega = \pi$).  In $t\in[0,T_\textrm{J}/2]$, $E_\textrm{MF}(t) = - E_{41} < 0$ increases with time as $\gamma_4(t)$ and $\gamma_1(t)$ becomes split, and the state is in $|\psi(t) \rangle \propto |0_{41} 0_{32} \rangle_t$.
In $t \in [T_\text{J}/2,3T_\text{J}/2]$, the state energy is zero, $E_\textrm{MF}(t)=0$, although $\gamma_3(t)$ and $\gamma_4(t)$ fuse or split in this time interval. It is because the state is in an equal-probability superposition of $|0_{43} 0_{21} \rangle_t$ and $|1_{43} 1_{21} \rangle_t$ [see Eq.~\eqref{halfperiod}]; the energy of $|0_{43} 0_{21} \rangle_t$, $-E_{43}(t)$, is cancelled by that of $|1_{43} 1_{21} \rangle_t$, $E_{43}(t)$. 
In $t\in[3T_J/2,2T_J]$, $E_\textrm{MF}(t) = - E_{41} < 0$ decreases with time, since $\gamma_4(t)$ and $\gamma_1(t)$ becomes fused and the state is in $|\psi(t)\rangle\propto|0_{41}0_{32}\rangle_t$.
At $t=2T_\textrm{J}$ the state returns to the initial state.
The other cases~\cite{otherEX} in Fig.~\ref{Fig:FJ} are understood similarly.

{\it $2n\pi$ Fractional AC Josephson effect.--} The $n T_\textrm{J}$ periodicity of the non-Abelian evolution is detected by the Josephson current of the junction.  In Fig.~\ref{Fig:FJ}(b) the current $I_\textrm{MF}$ mediated by the state $|\psi (t) \rangle$ of the four Majorana fermions $\gamma_k$'s follows~\cite{ChoiSimPRB} $I_\textrm{MF}=V_\textrm{DC}^{-1} dE_\textrm{MF}/dt$.
It has the period $nT_\textrm{J}$ and fractional Josephson frequency $2eV_\textrm{DC}/(nh)$, although the Hamiltonian $H(t)$ is $T_\textrm{J}$ periodic.
The total Josephson current $I_\textrm{J} = I_\textrm{MF} + I_\textrm{mid}$ that includes the contribution $I_\textrm{mid}$ from
the other midgap states is also $nT_\textrm{J}$ periodic, showing a $2n\pi$ fractional AC Josephson effect ($I_\textrm{J}$ is computed in Ref.~\cite{ChoiSimPRB}); $I_\textrm{mid}$ has the conventional period $T_\textrm{J}$.
This is also found for odd-parity states. 

The current $I_\textrm{MF}$ 
(or the Fourier components of $I_\textrm{J}$ with period $\ge 2T_\textrm{J}$) 
carries information about the evolution of a state formed by $\gamma_k$'s.
$I_\textrm{MF} \ne 0$ implies that the state gains or losses energy by fusion or splitting of $\gamma_k$'s.
When $I_\textrm{MF}=0$ in a time interval, $\gamma_k$'s are MZMs or the state is in  an equal-probability superposition of two temporal eigenstates of $H_\textrm{MF}(t)$ having the same parity (such as $|0_{43}0_{21}\rangle_t$ and $|1_{43}1_{21}\rangle_t$); the two eigenstates are particle-hole symmetry partners, mediating the equal amount of current in the opposite direction.
 
The period-elongation factor $n \ge 2$ is tuned by the voltage $V_\textrm{DC}$, since the state evolution depends on the dynamical-phase gain $\phi_\pm$ by fusion and splitting of $\gamma_k$'s.
This unique feature of our $2 n \pi$ effect,  the dependence of $n$ on $V_\text{DC}$, 
is absent in the known fractional Josephson effects~\cite{Clarke,ZhangKane,Orth,Klinovaja,Peng,Zazunov}.
It can be observed in time-resolved detection of the Josephson current $I_\textrm{J}$.
 
This can be also seen by measuring Shapiro spikes with applying an AC voltage  $V(t) = V_\textrm{DC} + V_\textrm{AC} \sin (\omega t)$ of frequency $\omega$ across the junction.
For $2eV_\text{AC} \ll \textrm{min} \{ 2e V_\textrm{DC}, \hbar \omega \}$, where the state evolution is affected weakly by $V_\textrm{AC}$, 
we numerically compute the evolution of the initial state $|0_{41} 0_{32} \rangle_0$ and 
the average of the Josephson current mediated by the state over a long time $\sim 10^4 T_\textrm{J}$ (Fig.~\ref{Fig:FJ}).
The average shows a peak (Shapiro spike) when $2e V_\textrm{DC} = ( q n / p) \hbar \omega$ is satisfied with $q = 1,2, \cdots$ and certain integers $p \ge 1$.
Only the spikes of $q=1$ are visible at $2eV_\textrm{AC} \ll \hbar \omega$, as the spike heights $\propto (\frac{2eV_\textrm{AC}}{ \hbar \omega})^q$~\cite{Tinkham}.
A spike at $2e V_\textrm{DC}/(\hbar \omega) = n / p$ implies the Fourier components $p$ of the $n T_\textrm{J}$-periodic supercurrent $I_\text{MF}(t)$ at $V_\textrm{AC} = 0$, expressed as $I_\text{MF}(t) = \sum_{p=1}^{\infty} a_{p} \sin\frac{2\pi p t}{nT_\text{J}}$; the spike height is proportional to $a_{p}$.
For example, in the $n=2, 3$ cases of Fig.~\ref{Fig:FJ}(a,b), the spikes appear at $2e V_\textrm{DC} / (\hbar \omega) = n/1, n/2, \cdots$.
In the $n=4$ case, the spikes appear at $2e V_\textrm{DC} / (\hbar \omega) =n/p = 4/1, 4/3, 4/5, \cdots$, since
$I_\text{MF}(t)$ does not have even-$p$ components in Fig.~\ref{Fig:FJ}(c) ($a_{2} = a_{4} = \cdots = 0$).
The $p \ge 2$ (high harmonics) components reflect the nonsinusoidal supercurrent resulting from the fusion and splitting of $\gamma_k$'s.
The spike positions are different from the case of conventional $2 \pi$ currents where a visible spike occurs only at $2 eV_\textrm{DC}/(\hbar \omega) = 1$ for  $2e V_\textrm{AC} \ll \hbar \omega$ (or  at $1,2,3,\cdots$ for larger  $V_\textrm{AC}$).
Notably, the sequence of the spike positions $n/p$ depends on $V_\textrm{DC}$ in our $2n \pi$ current. 
Detection of these features are within reach, as the Shapiro steps for $4 \pi$ supercurrents have been observed~\cite{Furdyna, Wiedenmann,Deacon,Laroche}.
Spikes at larger $V_\textrm{AC}$ is studied in Ref.~\cite{ChoiSimPRB}.

{\it Discussion.---} To have the non-Abelian evolution, the adiabatic regime $T_J \gg \hbar/E_0$ is required. We estimate~\cite{ChoiSimPRB} $\hbar/E_0 \sim 0.004 \,\text{ns}$, based on the minimum excitation energy $E_0 \sim \pi \hbar v_\textrm{arc} / (2l)$ of the setup with $2l \sim 2.5\,\mu\text{m}$ and $v_\textrm{arc} \sim v_F$. 
The regime is satisfied by $T_\text{J}=0.03 \sim 0.08 \,\text{ns}$ (equivalently, $V_\text{DC}= 25 \sim 74 \,\mu\text{V}$ and $\phi_+ = \pi/2 \sim 3\pi/2$). This time scale is shorter than $\hbar/k_B \mathcal{T} \sim 0.16 \, \text{ns}$ at temperature $\mathcal{T}=50$ mK
and quasiparticle poisoning time $0.1 \sim 1 \,\mu\text{s}$~\cite{Loss}. 
The parameters in Fig.~\ref{Fig:energy} satisfy $W > \xi$, with which $\gamma_k$'s form ``nonlocal'' electrons spatially separated by distance ($\sim W$) larger than $\xi$, suppressing state flip $|0_{41}0_{32}\rangle_0 \leftrightarrow |1_{41}1_{32}\rangle_0$ by Cooper-pair transfer between $\gamma_k$'s and the SCs.


According to our numerical study~\cite{ChoiSimPRB}, 
the $2n\pi$ supercurrent $I_\textrm{MF}$ is comparable with the $2\pi$ supercurrent $I_\textrm{mid}$ by the  midgap states,
and the total current $I_\textrm{J} = I_\textrm{MF} + I_\textrm{mid}$ shows the $2n \pi$ features of the Shapiro spikes in Fig.~\ref{Fig:FJ}; $I_\text{mid}$ leads to only one additional spike at $2eV_\textrm{DC} / (\hbar \omega) = 1$.
The $2n \pi$ fractional AC Josephson effect 
does not require fine tuning of the magnetic flux, $V_\textrm{DC}$, and the chemical potential of the TI~\cite{ChoiSimPRB}.
It also occurs, with modifications, in the presence of vortices in the SCs ($M_{a = \textrm{u,l}}\ne0$)~\cite{ChoiSimPRB}.
Our proposal is realizable with various effective $p$-wave SCs~\cite{Sophie,Harlingen,Fornieri,Ren,Wang, Ojanen,Bernevig,Tewari, Kim, Fujimoto,Shen}.

In summary, the $2n\pi$ fractional AC Josephson effect, a signature of the non-Abelian statistics, can occur in a {\it single} Josephson junction, a setup much simpler than the existing proposals. 
The junction can be used to demonstrate other non-Abelian statistics effects such as the non-commutativity of non-Abelian evolution $U$'s~\cite{ChoiSimPRB}.

We thank Myung-Ho Bae, Sungjae Cho, Gleb Finkelstein, David Goldhaber-Gordon, Gil-Ho Lee, Bernard Pla\c cais, Felix von Oppen, Yuval Oreg, Leonid Rokhinson, Ady Stern, and Bj\"orn Trauzettel for valuable discussions.  We are supported by Korea NRF (Grant No. 2016R1A5A1008184).

\end{document}